\documentclass[11pt]{article}
\usepackage{amsmath,amssymb,latexsym}

\begin{document}
\providecommand{\bo}{\boldsymbol}
\providecommand{\ra}{\rightarrow}
\def\Z4{\mathbb{Z}_4}
\newtheorem{theorem}{Theorem}[section]
\newtheorem{lemma}{Lemma}[section]
\newtheorem{corollary}{Corollary}[section]
\newtheorem{proposition}{Proposition}[section]
\providecommand{\un}{\underline}
\newcommand{\be} {\begin{equation}}
\newcommand{\ee} {\end{equation}}
\newtheorem{definition}{Definition}[section]
\newtheorem{example}{Example}[section]
\title{A Database of $\mathbb{Z}_4$ Codes}

\author{{Nuh~Aydin~and~ Tsvetan~Asamov}\\
        {Department~of~ Mathematics,~ Kenyon~ College,}\\
        {~ Gambier,~ OH~ 43022~USA}}

\maketitle

\begin{abstract}
There has been much research on codes over $\mathbb{Z}_4$,
sometimes called quaternary codes, for over a decade. Yet,  no
database is available for  best known quaternary codes. This work
introduces a new database for quaternary codes. It also presents a
new search algorithm called genetic code search (GCS), as well as
new quaternary codes obtained by existing and new search methods.

{\bf Keywords:} Quaternary codes, binary codes, cyclic codes,
quasi-cyclic codes, genetic code search.
\end{abstract}

\section{Introduction}

One of the main problems of coding theory is to construct codes
with best possible parameters. There are databases of best known
codes over small finite fields. For many years the online table
\cite{server1} has been the primary source of the records of the
best known codes over small fields. Recently, it is announced that
this table is discontinued due to the existence of  \cite{server2}
which often has more explicit information on constructions. The
computer algebra system MAGMA \cite{magma} has such a database
too. Moreover, a table of best known binary non-linear codes is
available at \cite{server3}.

For over a decade there has been intensive research on codes over
 $\mathbb{Z}_4$, integers modulo 4, sometimes called quaternary
codes. The term ``quaternary code" has been used both for codes
over $\mathbb{Z}_4$ and for codes over $\mathbb{F}_4$, the finite
field with 4 elements. In this paper, we shall use the term
exclusively for  $\mathbb{Z}_4$  codes. Among other results, some
good quaternary codes have been constructed
\cite{newcode},\cite{Z4},\cite{evenQC}, and \cite{cyclicZ4}.
Self-dual codes over $\Z4$ of length up to 9 are classified in
\cite{conway93}, and this is extended to length 15 in
\cite{fields98} (16 for Type II codes in \cite{plf97}). Rains has
classified optimal self-dual codes over $\Z4$ in \cite{rains99}. A
large number of self-orthogonal quasi-twisted (QT) $\Z4$ codes
have also been constructed \cite{glynn07}. Despite all this
research,  no database of best known quaternary codes is
available. The development of such a table has been started in
\cite{cyclicZ4}. We now have compiled a database of quaternary
codes. It is available at http://Z4Codes.info/ and it is being
updated continually. Unlike the tables at \cite{server2}, we do
not overwrite the existing entries when they are improved but
rather keep both the old and the new results. This strategy is
chosen primarily based on the fact that several different metrics
on quaternary codes, i.e. Hamming, Euclidean and Lee distance,
have been considered by the researchers. Moreover, for the sake of
easier communication, we have decided to provide researchers with
administrative privileges that would allow them to add their new
results to the table, as well as edit the existing entries.
Accounts can
be acquired by contacting the database editors via email.

In addition to the creation of the database, we have also devised
and implemented some search algorithms to find new quaternary
codes. The paper also describes and reports the results of these
searches. One of the search methods we consider in this paper is
the further exploration of the class of quasi-cyclic codes, which
has been the source of many of the new codes discovered in recent
years. The other method is called the ``progressive dimension
growth" (PDG) which is introduced recently in \cite{pdg} for
fields. In our work, it is modified for the ring $\Z4$ and the Lee
metric. Finally, extending some of the ideas behind PDG, we
implemented a new algorithm called ``genetic code search" (GCS)
that has produced better results than PDG over $\Z4$. The
following sections give more information about the algorithms. We
have used MAGMA for all computations.

\section{Basic Facts on Quaternary Codes}

A  {\em code} $C$ of length $n$ over $\Z4$ is a subset of
$\Z4^{n}$. $C$ is a {\em linear code} over $\Z4$ if it is an
additive subgroup of $\Z4^{n}$, hence a submodule of $\Z4^{n}$. In
this paper we will consider only linear codes over $\Z4$. An
element of $C$ is called a {\em codeword} and a {\em generator
matrix} is a matrix whose rows generate $C$. The {\em Hamming
weight} $w_H(x)$ of a vector $x=(x_1,x_2,\ldots,x_n)$ in $\Z4^n$
is  $|\{i: x_i\ne0\}|$. The {\em Lee weight} $w_L(x)$ of a vector
$x$ is $\sum_{i=1}^n \min\{|x_i|,|4-x_i|\}$.

The Hamming and Lee distances $d_H(x,y)$ and $d_L(x,y)$ between
two vectors $x$ and $y$ are $w_H(x-y)$ and $w_L(x-y)$,
respectively. The minimum Hamming and Lee weights, $d_H$ and
$d_L$, of $C$ are the smallest Hamming and Lee weights,
respectively, amongst all non-zero codewords of $C$.

The {\em Gray map} $\phi : \Z4^{n} \rightarrow \mathbb{Z}_2^{2n}$
is the coordinate-wise extension of the function from $\Z4$ to $
\mathbb{Z}_2^{2}$ defined by $0 \rightarrow (0,0), 1 \rightarrow
(1,0), 2 \rightarrow (1,1), 3 \rightarrow (0,1)$. The image,
$\phi(C)$, under the Gray map of a linear code $C$ over $\Z4$ of
length $n$ is a (in general non-linear) binary code of length
$2n$. The Gray map is an isometry from $(\Z4^{n},w_L)$ to
$(\mathbb{Z}_2^{2n},w_H)$. Therefore, the minimum Hamming weight
of $\phi(C)$ is equal to the minimum Lee weight of $C$.

Two codes are said to be {\em equivalent} if one can be obtained
from the other by permuting the coordinates and (if necessary)
changing the signs of certain coordinates. Codes differing by only
a permutation of coordinates are called {\em
permutation-equivalent}. Any linear code $C$ over $\Z4$ is
permutation-equivalent to a code with generator matrix $G$ of the
form \be \label{eqn:1} G = \left[
\begin{array}{ccc}
I_{k_1} & A_1 & B_1 + 2 B_2 \\
0 & 2 I_{k_2} & 2 A_2
\end{array}
\right],
\end{equation}
where $A_1, A_2, B_1$, and  $B_2$ are matrices with entries 0 or 1
and $I_k$ is the identity matrix of order $k$. Such a code has
size $4^{k_1}2^{k_2}$. The code is a free module if and only if
$k_2=0$. If $C$ has length $n$ and minimum Lee weight $d_L$, then
it is referred to as an $[n,4^{k_1}2^{k_2},d_L]$-code.

\subsection{Cyclic Codes over $\Z4$}

A cyclic  code over $\Z4$ is a $\Z4$-linear code which is
invariant under cyclic shifts where  the cyclic shift of an
$m$-tuple $(x_0,x_1, \ldots,x_{m-1})$ over $\Z4$ is the $m$-tuple
$(x_{m-1},x_0, \ldots, x_{m-2})$. Similarly to the case of finite
fields, cyclic codes over $\Z4$ of length $n$ are ideals in the
ring $\frac{{\mathbb{Z}}_4[x]}{\langle x^n-1\rangle}$  under the
usual identification of vectors with polynomials. Although
algebraically cyclic codes have the same structure over fields and
over $\Z4$ (ideals in a factor ring), the fact that $\Z4[x]$ is
not a unique factorization domain makes it more challenging to
find all cyclic codes over $\Z4$. For instance, computer algebra
systems (such as Magma and Maple), cannot directly provide
factorizations of $x^n-1$ for an arbitrary $n$. When $n$ is odd,
it is easier to obtain a factorization of $x^n-1$ and hence to
find all cyclic codes of length $n$. For an even $n$, the
situation is much harder. In fact, the factorization is not unique
in that case. In this paper we consider only cyclic codes of odd
length over $\Z4$. For the case of odd $n$ some of the most
important facts about ideals of the relevant ring and the
factorization of $x^n-1$ are summarized below, and they can be
found in \cite{pless}, \cite{vanlint} or \cite{wan97}. For the
case of even $n$, we refer the reader to
\cite{reversible},\cite{gencyclic}, and \cite{oddlyeven}.

For an odd positive integer  $n$, $x^n-1$ can be factored into a
product of finitely many pairwise coprime basic irreducible
polynomials over ${\mathbb{Z}}_4$. Also, this factorization is
unique up to ordering of the factors \cite{pless,wan97}. In fact,
we have the following: if $f_2(x)|(x^n-1)$ in ${\mathbb{Z}}_2[x]$
then there is a unique, monic polynomial $f(x)\in
{\mathbb{Z}}_4[x]$ such that $f(x)|(x^n-1)$ in ${\mathbb{Z}}_4[x]$
and $\overline{f(x)}=f_2(x)$, where $\overline{f(x)}$ denotes the
reduction of $f(x)$ modulo 2 \cite{wan97}. The polynomial $f(x)$
is called the Hensel lift of $f_2(x)$. There are well-known
methods of finding this polynomial, such as Graeffe's method
\cite{wan97}. Therefore, there is a one-to-one correspondence
between irreducible factors of $x^n-1$ over ${\mathbb{Z}}_2$ and
irreducible factors of $x^n-1$ over ${\mathbb{Z}}_4$.

 Once the factorization of $x^n-1$ over
${\mathbb{Z}}_4$ is obtained, the ideals of
$R:=\frac{{\mathbb{Z}}_4[x]}{\langle x^n-1\rangle}$ can be
determined. For an odd positive integer $n$, any ideal $I$ of the
ring $R$ has a generator of the form a $I=\langle f(x)h(x),
2f(x)g(x)\rangle $ where $f(x)g(x)h(x)=x^n-1$ \cite{pless,wan97}.
Moreover, $|I|=4^{\deg g(x)}2^{\deg h(x)}$. It follows that the
number of cyclic codes of length $n$ is $3^r$, where $r$ is the
number of irreducible factors of $x^n-1$ \cite{pless}.

Finally, it can be shown that any ideal of $R$, for an odd $n$, is
a principle ideal, with a generator of the form
$p(x)=f(x)h(x)+2f(x)$ (or equivalently $p(x)=f(x)h(x)+2f(x)g(x)$)
where $f(x), g(x), h(x)$ are as above \cite{pless,wan97}.

{\bf Remark 1} When $x^n-1$ has $r$ irreducible factors over a
field, the total number of cyclic codes is $2^r$. We have a larger
number over $\Z4$ due to the existence of non-free codes (over a
field all codes are free).

{\bf Remark 2} The generator polynomial $p(x)$ of an ideal of $R$
described above does not necessarily divide $x^n-1$. For example,
let $n=3, f(x)=1$, and $h(x)=x-1$, then $p(x)=x+1$ and
$p(x)\not|{(x^3-1)}$. When $h(x)=1$, $p(x)=3f(x)=-f(x)$ does
divide $x^n-1$. It is shown in \cite{Z4} that the cyclic code
generated by $p(x)$ is a free module if and only if $p(x)$ divides
$x^n-1$.

\subsection{Quasi-Cyclic Codes over $\Z4$}

Much research has focused on the class of quasi-cyclic (QC) and
the related class of quasi-twisted (QT)  codes, and many new codes
over small finite fields have been discovered within these
classes. Some of these results can be found in
\cite{main},\cite{daskalov1,daskalov2},\cite{chain} and
\cite{siap}. In addition to the case of a field, QC codes over
rings, especially over $\Z4,$ have been studied as well. QC codes
over $\Z4$ are first studied in \cite{Z4}, where a number of
``good" quaternary codes are obtained. A quaternary linear code
$C$ with parameters $[n,4^{k_1}2^{k_2},d]$ (where $d$ is the Lee
weight) is called {\it good} if $d>d'$, where $d'$ is the minimum
distance of a best known binary linear code of length $2n$, and
dimension $2k_1+k_2$, i.e. if the Gray image of $C$ has a larger
minimum distance than the comparable binary linear code.
Similarly, a quaternary code will be called {\it decent} if its
Gray image has the same parameters as the best known binary code
(i.e., if $d=d'$). The reason for this type of comparison is that
even though the Gray image of a quaternary code is most likely
non-linear, we do not have any other means of testing how good the
parameters of a $\Z4$ code are due to the
facts that\\
 a) the table \cite{server3} is much smaller than the
tables for binary linear codes (it only goes up to minimum
distance 29), and \\
b) there are no extensive tables of quaternary codes.\\ We believe
that the table presented in this paper will meet a need in this
area. The researchers are welcome to report and enter their codes
to this database.

 Next, we summarize some of the basic facts concerning the
structures of QC codes. A more detailed treatment can be found in
\cite{main} for QC codes over fields, and in \cite{Z4} for QC
codes over $\Z4$. A linear code over a ring is called $l-$QC if it
is invariant under the cyclic shift by $l$ positions.
Algebraically, an $l$-QC code of length $n=ml$ over a ring $R$ can
be viewed as an $R[x]/\langle x^m-1\rangle$ submodule of
$(R[x]/\langle x^m-1\rangle)^l$. Then an $r$-generator QC code is
spanned by $r$ elements of $(R[x]/\langle x^m-1\rangle)^l$. In
this paper, as is the case in most of the literature, we restrict
ourselves to $1$-generator QC codes. The following is a
generalization of an important result about 1-generator QC codes
\cite{main}, \cite{Z4} that has been used in many of the recent
work \cite{daskalov1},\cite{daskalov2}. The ring $R$ can be a
finite field, or $\Z4$.

\begin{theorem}\label{qcbound}
Let $C$ be a $1$-generator $l$-QC code of length $n=ml$ with a
generator of the form:
\begin{equation}\label{eq:qcbound}
\bo{g(x)}=(f_1(x)g(x),f_2(x)g(x),\ldots ,f_l(x)g(x))
\end{equation}
where $g(x)|(x^m-1), g(x),f_i(x)\in R[x]/\langle x^m-1\rangle,$
and $(f_i(x),h(x))=1$, $h(x)=\frac{x^m-1}{g(x)}$ for all $ 1\leq
i\leq l$. Then  $l\cdot d \leq d(C)$, where $d$ is the minimum
distance of the cyclic code generated by $g(x)$, and $d(C)$ is the
minimum distance of $C$. Moreover, the dimension of $C$ is equal
to the dimension of the cyclic code generated by $g(x)$.
\end{theorem}

In terms of generator matrices, the QC codes can be characterized
as follows.

Let

\begin{equation}\label{twistulant}
G_0=\left[
\begin{array}{ccccc}

g_0 & g_1 & g_2& \ldots & g_{m-1} \\
g_{m-1} & g_0 &g_1 & \ldots  & g_{m-2}\\
g_{m-2} & g_{m-1} & g_0 & \ldots & g_{m-3}\\
\vdots & \vdots & \vdots &   & \vdots \\
g_{1} & g_{2} & g_{3} &\ldots & g_{0}
 \end{array}
\right]_{m\times m}
\end{equation}

An  $(m\times m)$ matrix of the type $G_0$ is called a circulant
matrix of order $m$ or simply a circulant matrix.

It is well-known that  the generator matrices of $QC$ codes can be
transformed into blocks of circulant  matrices by a suitable
permutation of columns. Therefore, any 1-generator QC code is
permutation equivalent to a linear code generated by a matrix of
the form

\begin{equation*}
\left[
\begin{array}{cccc}
G_{1} & G_{2} & \ldots & G_{l} \\
 \end{array}
 \right]_{m\times n}
\end{equation*}
where each  $G_{k}$ is a circulant matrix of the form (3).

\subsection{A New QC $\Z4$ Code}

Based on the search results for new codes over fields and over
$\Z4$, it is natural to search for new quaternary codes in the
class of QC codes over $\Z4$. Although this kind of search is
carried out in \cite{cyclicZ4},\cite{Z4}, and \cite{evenQC}, a
more complete search is still possible. For each odd  integer $m$
up to length 63, we produced all cyclic codes i.e., their
generators, $p(x)$,(free or non-free), based on the results
described in section II-A. We then searched for new QC codes of
the form $(p(x),p(x)f_1(x),\dots,p(x)f_{l-1})$. In most cases we
used $l=2$ (in a few cases we also let $l=3,4$). Our search
revealed a good (and new) quaternary QC code with parameters
$[86,4^{15}2^0,55]$, whose Gray image (which is non-linear) is a
binary $(172,2^{30},55)$-code. The best known binary linear code
of length 172, and dimension 30 has minimum distance 54. The
generators and the Lee weight enumerator of this code are as
follows:\\
 $g(x)=x^{15}+3x^{14}+2x^{13}+3x^{12}+2x^9+2x^8+2x^7+2x^6+x^3+2x^2+x+3$,\\
$f(x)=x^{28}+x^{27}+3x^{26}+2x^{25}+x^{24}+2x^{22}+3x^{21}+x^{20}+
3x^{19}+2x^{18}+x^{17}+x^{16}+2x^{15}+3x^{14}+2x^{13}+x^{12}+x^{11}
+2x^{10}+3x^9+x^8+3x^7+2x^6+x^4+2x^3+3x^2+x+1$

\noindent $h(x) = 1$ so that $g(x)f(x)h(x)=x^{43}-1$.

\noindent Let $p(x)= f(x)h(x)+2f(x)$, then  the polynomial $p(x)$
generates a free quaternary cyclic code with parameters
$[43,4^{15}2^0,16]$. The search revealed that with the choice of
$f_1 = 2x^{13}+x^{12}+x^{10}+2x^9+3x^8+x^7+
3x^6+3x^5+3x^4+2x^2+x$, the 1-generator QC code generated by
$(p(x),p(x)f_1(x))$ has parameters $[86,4^{15}2^0,55]$. Its Lee
weight enumerator is given below where the bases are weights, and
exponents are number of codewords of that weight

\noindent $0^1 55^{774} 56^{1591} 57^{3698} 58^{5289} 59^{13244}
60^{24639} 61^{43602}
 62^{74691}\\
 63^{132870} 64^{233877} 65^{374100} 66^{614169}
67^{970854} 68^{1502291}\\
 69^{2252598} 70^{3320202} 71^{4791318}  72^{6689811} 73^{9186262}
 74^{12274866}\\
  75^{15998236} 76^{20463442} 77^{25598416}
  78^{31106974}
 79^{36948696} 80^{43080625}\\
  81^{48872424} 82^{54121520} 83^{58775152} 84^{62257851} 85^{64430426}
 86^{65299285}\\
  87^{64550138} 88^{62322437} 89^{58728454} 90^{54154888} 91^{48850752}
  92^{42923718}\\
 93^{37050520} 94^{31176720} 95^{25516630} 96^{20478707} 97^{16029368}
 98^{12290346}\\
  99^{9187466} 100^{6707312} 101^{4753392} 102^{3279137} 103^{2255178}
 104^{498636}\\
  105^{982292} 106^{634379}  107^{382872} 108^{227341} 109^{134590}
 110^{76067}\\
 111^{41452} 112^{21930} 113^{10578} 114^{6665}
115^{3440} 116^{1118} 117^{1032} 118^{172}\\ 120^{129} 121^{86}
122^{86} 129^2$

\section{Quaternary Codes from Inverse Gray Map}

The Gray map is usually used to obtain binary codes (usually
non-linear) from quaternary codes (usually linear). However, we
can also use its inverse to obtain quaternary codes (most likely
non-linear) from a given binary code. If we take a binary code
with parameters $[2n,2k,d]$ then the inverse Gray map yields a
quaternary code (which is most likely to be non-linear) with
parameters $(n,4^k,d)$. Taking advantage of existing databases for
binary linear codes, we considered quaternary codes obtained this
way from best known binary codes. This method contributed
thousands of (non-linear) codes to the database.

\section{Genetic Code Search}

It is well known that computing minimum distance of an arbitrary
linear code is an NP-hard problem \cite{vardy}. This result gives
an insight about why there does not exist an efficient, general
purpose search algorithm to find good linear codes. All known
search methods/algorithms for linear codes work well in some
special cases. Recently, a new search algorithm has been
introduced  that works well for most parameter ranges over small
fields \cite{pdg}. In a large number of cases the algorithm
produced linear codes with best known parameters, and in several
cases generated new codes (``record breakers"). In our work we
adopted this algorithm for the ring $\Z4$ and the Lee metric. We
refer the reader to \cite{pdg} for  further details. Originally,
as implemented for the field case, PDG did not work very well for
$\Z4$. Therefore, we introduced changes inspired by genetic
algorithms. \textbf{We start off with an empty code and gradually expand
it. At each step we examine multiple mutations of a single generator matrix. 
Thus GCS is not a typical genetic algorithm, in the sense that it operates on a 
 single element and crossover between generator matrices is not considered.} Here we present the details of GCS for free codes, i.e. $K=K_1,
K_2 = 0$. 

\begin{center}
\begin{tabular}{l}
\hline \textbf{Initialize} Set the input parameters and
initialize\\
sets and variables. \\
\hline \ $\mbox{Set } \textbf{N,K,T};$ \\
\ $ \mbox{Use the binary record table to determine } D;$\\
\ $BitShifts = \{1,2,3\};$\\
\ $S = \{ \};$\\
\ $G = [0]; t = 1; k = 1;$\\
\hline
\end{tabular}\end{center}
The length $N$ and dimension $K$ are the two input parameters. \textbf{Based on 
those we can determine a desired Lee minimum distance $D$ using the record table for binary codes.} In addition, we also specify a small integer $T$, $1\leq T\leq N-K$.
Greater $T$ implies a better chance for the construction of a good
code but that benefit comes at the expense of increased
computational time. \textbf{The output of the algorithm is a linear code over $Z_4$ 
of length $N$ and minimum Lee distance $d\geq D$.}
 
\begin{center}
\begin{tabular}{l}
\hline
\textbf{General GCS Algorith} \\
\hline \ Initialize \\
\ \textbf{while} $((t\leq T)$ and $(k\leq K))$ \textbf{do}\\
\   \ $S = S \cup\{ (K + kN),\dots,((k+1)N - 1)\};$\\
\   \ $G_{old} = [0];$ $G_{new} = [0];$\\
\   \ $G_{temp} = [0];$ $G_{old}[k][k] = 1; $\\
\   \ $d = 1;$\\
\   \ Search for a suitable matrix $G_{old}$\\
\   \ \textbf{if} $(d\geq D)$ \textbf{then}\\
\   \   \ $G = G + G_{old};$\\
\   \   \ $k = k+1;$\\
\   \ \textbf{end if}\\
\ \textbf{end while}\\

\hline
\end{tabular}\end{center}
The search for a suitable mutation matrix $G_{old}$ presents the
heaviest computational task. As the size of the set $S$ increases
with successive dimensions, so does the number of possible
mutation matrices. This is a key difference between PDG and GCS.
The process terminates either when a suitable mutation matrix is
found or the specified level of $T$ is reached.
\begin{center}
\begin{tabular}{l}
\hline \textbf{Search for a suitable matrix }$G_{old}$ \\
\hline
\ \textbf{while} $((t\leq T)$ and $(d < D))$ \textbf{do}\\
\   \ $increment\_t = true$; \\
\   \ $RedundancyShifts = \mbox{The set of all subsequences}$\\
\   \ of $BitShifts$ of length $t$;\\
\   \ $Positions = $ The set of all subsets of $S$ of size $t$;\\
\   \ \textbf{for} $p$ in $Positions$ \textbf{do}\\
\   \   \ $G_{new} = G_{old};$\\
\   \   \ \textbf{for} $r$ in $RedundancyShifts$ \textbf{do}\\
\   \   \   \ \textbf{for} $i = 1$ to $t$ \textbf{do}\\
\   \   \   \   \ $G_{new}[p(i) \div N][(p(i) \mod N) + 1] =$\\
\   \   \   \   \ $=G_{old}[p(i) \div N][(p(i) \mod N) + 1] +
r(i)$;\\
\   \   \   \ \textbf{end for}\\
\   \   \   \ $G_1 =G + G_{new};$ $C = <G_1>;$\\
\   \   \   \   \textbf{if} $(MinimumLeeWeight(C)>d)$ \textbf{then}\\
\   \   \   \   \   \ $d=MinimumLeeWeight(C);$\\
\   \   \   \   \   \ $G_{temp} = G_{new};$\\
\   \   \   \   \   \ $increment\_t = false$;\\
\   \   \   \   \   \ \textbf{break} $p$;\\
\   \   \   \   \textbf{end if}\\
\    \   \ \textbf{end for}\\
\   \ \textbf{end for}\\
\   \ $G_{old} = G_{temp};$\\
\   \ \textbf{if} $increment\_t$ \textbf{then}\\
\   \   \ $t = t+1;$\\
\   \ \textbf{end if}\\
\ \textbf{end while}\\
\hline
\end{tabular}
\end{center}

Below is a table of small quaternary codes obtained with GCS whose
Lee distances are equal to the minimum Hamming distances of the
corresponding binary linear codes. We call such a code ``decent".
This is significant, considering that all binary linear codes up
to length 32 are optimal. Based on our experience and the results
from the literature, constructing decent codes is a challenging
task. Besides, many of the decent $\Z4$ codes may very well be
regarded new.
\begin{center}
\begin{tabular}{l}
\hline Free decent quaternary linear codes.\\
\hline\\
$N$\    \   \ $K_1$ \\
\hline
$10:$ $1,2,3,4,5,6,7,8,9$\\
$11:$ $1,3,4,7,8,9,10$\\
$12:$ $1,2,3,4,5,8,9,10,11$\\
$13:$ $1,2,3,4,5,7,8,9,10,11,12,$\\
$14:$ $1,2,3,6,8,10,11,12,13$\\
$15:$ $1,2,3,4,7,9,11,12,13,14$\\
$16$ $1,2,6,7,8,9$\\
$17:$ $1,2,3,4,5$\\
$18:$ $1,2,3$\\
$19$ $1,2,3$\\
$20:$ $1,2,3,4$\\
$21:$ $1,2,3,11$\\
$27:$ $1,2,3$\\
\hline
\end{tabular}\end{center}

\section{Conclusion}

In this work, we introduce a new database of $\Z4$ codes that is
available online that can be conveniently updated by researchers.
The database has been populated using several different search
methods. We present a survey of some of the recent and promising
methods to find new quaternary codes. Search with one of these
methods has yielded a good quaternary code. We also introduce a
new search method that has yielded decent codes. We invite
researches to search for new quaternary codes using known methods
or devising new ones, and update the database with any new codes
discovered. There is much room for improvement on this database.

\end{document}